\documentclass{iopart} 
\usepackage{iopams,graphicx} 

\begin{document}

\title{Emergent Quantum Jumps in a Nano-Electro-Mechanical System}
%\shorttitle{Title} %Insert here a short version of the title if it exceeds 70 characters

\author{Kurt Jacobs$^{1,2}$ and Pavel Lougovski$^1$} 

\address{$^1$ Quantum Science and Technologies Group, Hearne Institute for Theoretical Physics, Louisiana State University, 202 Nicholson Hall, Tower Drive, Baton Rouge, LA 70803, USA}

\address{$^2$ Department of Physics, University of Massachusetts at Boston, 100 Morrissey Blvd, Boston, MA 02125, USA} 

\ead{kurt.jacobs@umb.edu}

\pacs{03.65.Ta, 85.85.+j, 85.35.Ds}

\begin{abstract}
We describe a nano-electromechnical system that exhibits the ``retroactive'' 
quantum jumps discovered by Mabuchi and Wiseman [Phys. Rev. Lett. 81, 4620 (1998)]. 
This system consists of a Cooper-pair box coupled to a nano-mechanical resonator, in which the latter is continuously monitored by a single-electron transistor or quantum point-contact. Further, we show that these kinds of jumps, and the jumps that emerge in a continuous quantum non-demolition measurement are one and the same phenomena. We also consider manipulating the jumps by applying feedback control to the Cooper-pair box.
\end{abstract}

\maketitle

In 1998 Mabuchi and Wiseman discovered the existence of emergent jumps in purely diffusive, measured cavity QED system, and the termed the phenomena ``retroactive'' quantum jumps~\cite{Mabuchi98}. Since that time it has also been discovered that quantum jumps will emerge in a noisy system subjected to a continuous quantum non-demolition (QND) measurement. Examples are the continuous measurement of the energy of a two-level system, such as a Cooper-Pair Box (CPB)~\cite{Goan01}, and the measurement of the energy of a nano-mechanical resonator~\cite{Santamore04, Santamore04b,Martin07,Jacobs07b,Buks06,Jacobs08}. Our main purpose here is to show that the jumps discovered by Mabuchi and Wiseman, and the jumps in a QND measurement are, in fact, the same phenomena having the same underlying origin.  While there was no explicit QND measurement in the cavity QED scenario of Mabuchi and Wiseman, the ``retroactive'' jumps are nevertheless due to a QND measurement that arises as a result of the interaction between two subsystems. This is a QND measurement on one subsystem mediated through the other subsystem. These jumps may therefore be described as arising from an {\em indirect} QND measurement, rather than a direct one. Thus, if one interprets the cQED jumps of Mabuchi and Wiseman as retroactive, then the jumps induced by all QND measurements, whether on quantum or classical systems, are retroactive in the same sense.  

In the process of our analysis will also do a number of other things. We first show how to design a nano-electro-mechanical system so that jumps of the ``retroactive'' type emerge in the diffusive motion. This system consists of a nano-mechical resonator linearly coupled to a CPB~\cite{Makhlin01}, where the position of the resonator is continuously measured. This measurement can be implemented using a single-electron transistor (SET)~\cite{Naik06} or a quantum point contact (QPC)~\cite{Korotkov}.  The dynamics of this system are not identical to the cavity-QED system analyzed in~\cite{Mabuchi98}, and this difference makes it easier to see the connection with the usual direct QND measurement. We also show that it is possible to effect a degree of control over the jumps by applying feedback control to the CPB. 

If one places a CPB adjacent to a nano-resonator, and places a voltage on the nano resonator, then the resonator feels a linear force from the CPB, and this depends upon which state the CPB is in. (That is, whether or not the CPB contains an excess Cooper-pair). The Hamiltonian describing the dynamics of the coupled systems is~\cite{Armour02} 
\begin{equation}
  H = \hbar [  \omega_{\mbox{\scriptsize R}} a^\dagger a + \lambda \sigma_z (a+a^\dagger) +  \omega_{\mbox{\scriptsize C}} \sigma_z + \omega_{\mbox{\scriptsize J}} \sigma_x ] .
  \label{Ham}
\end{equation} 
Here $a$ is the annihilation operator for the resonator, which has angular frequency $\omega_{\mbox{\scriptsize R}}$. The frequency corresponding to the charging energy of the CPB is $\omega_{\mbox{\scriptsize C}}$, and the Josephson tunneling frequency is $\omega_{\mbox{\scriptsize J}} \sigma_x$. The tunneling term allows us to perform $\sigma_x$ rotations using voltage pulses, but otherwise can be ignored since we set $\omega_{\mbox{\scriptsize C}} \gg \omega_{\mbox{\scriptsize J}}$. The interaction between the resonator
and the CPB is the linear force that the resonator feels from the charge on the CBP, and the strength of this force determines $\lambda$. The expression for the above rate parameters in terms of the physical configuration of the resonator and CPB may be found, for example, in~\cite{Armour02,Hopkins03}.

As indicated above, the position of the resonator can be monitored continuously using an SET or QPC. We model this measurement process as an inefficient, but otherwise ideal, continuous position measurement~\cite{JacobsSteck06,Hopkins03,Brun02}. The continuous stream of measurement results (often referred to as the measurement record) is $r(t)$ where
$dr = \langle x\rangle dt + dW/\sqrt{8\eta k}$ and $dW$ is an increment of Gaussian white noise~\cite{WienerIntroPaper}. Here $k$ is a measure of the rate at which information is extracted from the system, and which we will refer to as the measurement strength. The
parameter $\eta$ is called the efficiency of the measurement. The interaction of the system with its environment (including the SET) continually carries information away from the system (at a rate proportional to $k$), and $\eta$ gives the fraction of this information which is actually collected by the observer. The resulting dynamics of the system density matrix, $\rho$, is given by the Stochastic Master Equation (SME)~\cite{DJ,Korotkov}
\begin{equation}
  d\rho =    (-i/\hbar)[H,\rho] dt - k [x,[x,\rho]] dt \nonumber   + \sqrt{2 \eta k} (x \rho + \rho x - 2\langle x\rangle \rho ) dW 
  \label{msme}
\end{equation}
where $H$ is given by Eq.(\ref{Ham}) above and $x$ is the position
operator for the resonator. As such, $k$ has units of $\mbox{m}^{-2}
s^{-1}$, and so we define the corresponding dimensionless rate
$\tilde{k} = k (\hbar/2m\omega)$.

We will consider two modifications of the above basic dynamics. The
first is that we will modulate the interaction strength $\lambda$
between the resonator and the CPB at the resonant frequency of the
resonator, so that $\lambda = \lambda_0 \cos(\omega_{\mbox{\scriptsize
R}} t)$. This can be done by varying the voltage on the resonator. The
result is to allow the CPB to drive the resonator at its resonant
frequency, generating the maximum steady-state displacement of the
resonator. The second modification is the application of a real-time
feedback loop~\cite{Ruskov05,Hopkins03} to damp the motion of the
resonator. This means that the observer continually applies a force
$F(t) = -\gamma \langle p(t) \rangle$ to the resonator, where $\langle
p(t) \rangle = \mbox{Tr}[p\rho(t)]$ is the observers maximum
likelihood estimate of the momentum of the oscillator at each time
$t$. The result of this is to apply a (somewhat noisy) frictional
damping force to the resonator.

We must also include in the dynamics the effects of temperature on the
resonator. The resonator is in contact with a thermal bath that
induces damping and injects noise into the resonator. Since the
quality factor of the resonator is above $10^4$, the thermal damping
is much smaller than the damping that will be induced via the feedback
loop, and as a result we simply subsume this damping into the
feedback. The thermal noise can be taken into account by choosing an
appropriate value for the efficiency $\eta$~\cite{Hopkins03}. The
noise introduced by the (low temperature) thermal bath is just as if a
position detector was carrying information away at the rate
$k_{\mbox{\scriptsize therm}} = (m \omega_{\mbox{\scriptsize R}}
\Gamma)/(2 \hbar) \coth (\hbar \omega_{\mbox{\scriptsize R}})/(2 k_B
T)$~\cite{Caldiera89, Hopkins03}, where $\Gamma$ is the thermal
 damping rate and $T$ is the temperature. To include the
thermal noise one therefore replaces $k$ in Eq.(\ref{msme}) with
$k_{\mbox{\scriptsize tot}} = k+k_{\mbox{\scriptsize therm}}$, and
$\eta$ with $\eta_{\mbox{\scriptsize tot}} =
(k/k_{\mbox{\scriptsize tot}}) \eta$ where $\eta$ is the SET measurement efficiency. 

The final thing to include is the environmental noise on the CPB. The
kind of noise that is of interest to us is noise which causes
diffusion between the two energy eigenstates. It is this noise which,
coupled with the system's dynamics, induces the quantum jumps. Thermal
noise is of this type, and is usually modeled with the master equation
$\dot{\rho} = \kappa \{ (\xi + 1) {\cal D}[\sigma_-] + \xi {\cal
D}[\sigma_+] \} \rho $, where $\xi = 1/(e^{\hbar \omega_{\mbox{\tiny
C}}/(k_B T)} -1) $, and ${\cal D}[a]\rho \equiv [a^\dagger a,\rho]_+ -
2 a \rho a^\dagger$ for any operator $a$. However, the charging energy
of a CPB is usually chosen so that $\hbar \omega_{\mbox{\tiny C}}/(k_B
T) \ll 1$. In this case the thermal noise will only cause the upper
state to decay, rather than generate diffusion between the two.

Nevertheless there are other ways to induce the desired diffusion. One
is to apply a stochastic sequence of pulses to a voltage gate, where
the pulses bring the CPB to the degeneracy point. If the random pulse
lengths are short compared to $\omega_{\mbox{\scriptsize J}}$, then
the Josephson tunnelling term generates an evolution described by a
Hamiltonian $\zeta(t) \sigma_x$, where $\zeta(t)$ is white noise with
autocorrelation $\langle \zeta(t)\zeta(t+\tau) \rangle = \kappa
\delta(\tau)$. The result is the master equation $\dot{\rho} = \kappa
{\cal D}[\sigma_x] \rho$. This master equation closely emulates the
thermal master equation in the limit in which $\xi \gg
1$~\footnote{Making the approximation $\xi\gg 1$, the thermal noise
term becomes $k\xi (2\rho - \sigma_x \rho \sigma_x - \sigma_y \rho
\sigma_y)$. Since the system Hamiltonian is approximately symmetric in
$x$ and $y$, the effect on the system is equivalent to $2k\xi (\rho -
\sigma_x \rho \sigma_x)$.}.  Thirdly, one could instead increase the
Josephson term.  Due to the measurement dynamics to be discussed
later, this should have the same effect as thermal noise.

In our numerical simulations we choose to explicitly model the second
noise source discussed above, that of stochastic driving proportional
to $\sigma_x$ As a result, the full dynamics of the resonator-CPB
system is
\begin{eqnarray}
  \!\!\!\! d\rho & = &   (-i/\hbar)[H(t) - \gamma x\langle p\rangle,\rho] dt -k_{\mbox{\scriptsize tot}} [x,[x,\rho]] dt \nonumber   \nonumber \\
           & + & \kappa {\cal D}[\sigma_x] \rho dt  + \sqrt{2 \eta_{\mbox{\scriptsize tot}} k_{\mbox{\scriptsize tot}}} ( x \rho + \rho x - 2\langle x\rangle \rho ) dW 
\end{eqnarray}
Following the analysis of the cavity-QED system by Mabuchi and Wiseman, 
we now examine the
steady-state of the Hamiltonian $H(t)$ including the feedback damping,
when the CPB is in either of it's energy eigenstates. We will denote
these eigenstates by $|\!\pm\rangle$ --- they correspond to the
presence or absence of a Cooper-pair in the box.  In these two cases
the resonator has the effective Hamiltonian
\begin{equation}
  H_\pm = \hbar \omega_{\mbox{\scriptsize R}} a^\dagger a 
                 - (\gamma \langle p\rangle 
          \pm F \cos(\omega_{\mbox{\scriptsize R}} t)) x ,
    \label{HamRes}
\end{equation}
where $F = \lambda \sqrt{2\hbar m\omega_{\mbox{\scriptsize R}}}$ is
the maximum value of the driving force. Since the Hamiltonian is
linear, the dynamics of the expectation values of $x$ and $p$ are
simply those for the equivalent classical system, namely that of a
driven, damped harmonic oscillator. As a result, the steady-state
solution for $\langle x (t) \rangle$ is
\begin{equation}
  \langle x (t) \rangle = \frac{F}{\gamma \omega_{\mbox{\scriptsize R}} m} \cos(\phi + \omega t) \; , \;\; \phi = \mp \frac{\pi}{2} ,
  \label{ssx}
\end{equation}
where $m$ is the mass of the resonator. The state of the resonator in each case is a (somewhat squeezed) Gaussian state. Thus, like the cavity-QED system in~\cite{Mabuchi98}, our system has two stable orthogonal states, each of which is a product of a distinct state of the two-level system with a distinct Gaussian state of the resonator. In our case it is the steady-state phase of the resonator, $\phi$, that is tightly correlated with the orthogonal states of the CPB. 

The measurement of the position of the oscillator continually provides
the observer with information regarding the location of the oscillator
in phase space, and thus about the phase of the oscillations. Since
this phase is correlated with the eigenstates of the oscillator, the
measurement will tend to continually collapse the state of the CPB to
one of its eigenstates. Because of this it is only the two eigenstates
that are stable against the measurement process, and this is the
reason that the two steady-states given by Eq.(\ref{ssx}) are
important --- if the environmental noise is to induce jumps in the
system it will be between these two stable states. 

We now simulate the full dynamics of the observed
nano-electro-mechanical system, including the environmental noise. In
specifying values for the system parameters, we will quote all rate
constants in terms of the frequency of the resonator $f =
\omega_{\mbox{\scriptsize R}}/(2\pi)$. We set the interaction strength
$\lambda = 0.5 f$ and the feedback damping rate $\gamma = 0.25 f$,
both of which are easily achievable ~\cite{Armour02,Hopkins03}. We set
the SET measurement strength at $\tilde{k}= 0.01f$, which is also not
difficult to achieve, certainly with $f$ as high as $10
\mbox{MHz}$~\cite{Hopkins03}.  We find from our simulations that
relatively high efficiency ($\eta_{\mbox{\scriptsize tot}} \geq 0.7$)
is required for the observer to effectively track the quantum jumps.
This requires that the SET have high efficiency, and that
$k_{\mbox{\scriptsize therm}} \ll k$. The question is still open as to
whether such an efficiency can be reached with an
SET~\cite{Clerk04}, but it is estimated that a QPC should in theory be 
able to achieve efficiencies above $\eta=0.8$~\footnote{A. N. Korotokov, private
communication}. Setting $Q=10^5$, and using the parameters
in~\cite{Hopkins03}, gives $k_{\mbox{\scriptsize therm}} \approx 5
k$. Thus a factor of 20 increase in $k$ from that configuration would
be required. While we would expect this to be possible, the overall
efficiency requirement is the most challenging in the scenario.
Finally, we choose the CPB noise strength to be $\kappa = 0.01f$.

\begin{figure}
\begin{center}
\leavevmode\includegraphics[width=8.5cm]{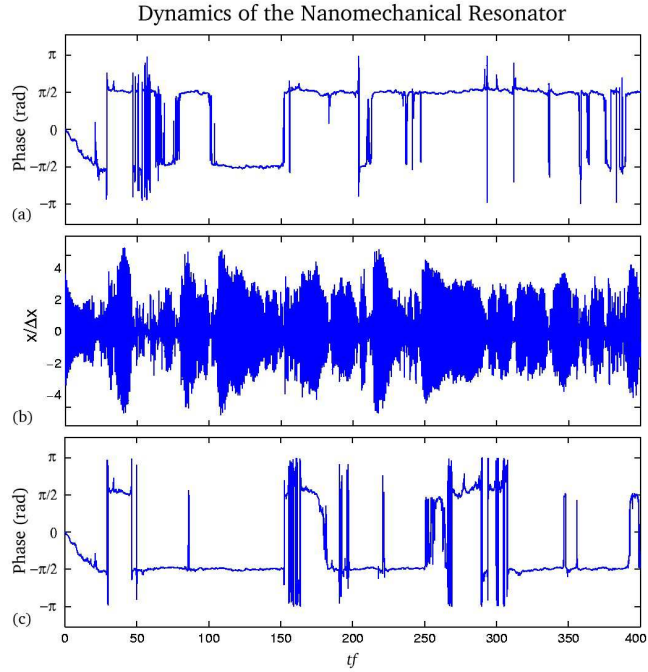}
\end{center}
\caption{Here we plot the evolution of the nanomechanical resonator under continual position measurement: (a) the phase of the resonator; (b) the mean position of the resonator; (c) the phase of the resonator when feedback control is applied to the Cooper-pair box. For (a) and (b) $\eta_{\mbox{\scriptsize tot}}=0.7$, and for (c) $\eta_{\mbox{\scriptsize tot}}=0.95$.}
\label{fig1}
\end{figure}

In Figure~\ref{fig1}(a) we show the evolution of the phase of the
resonator, $\theta(t)$, which we define by the relation
$A(t)e^{i\theta(t)-i\omega_{\mbox{\scriptsize R}}t} = \langle
\tilde{x}(t) \rangle + i\langle \tilde{p}(t) \rangle$. Here
$\tilde{x}$ is as defined above, and $\tilde{p} = -i(a-a^\dagger)$.
We start the resonator in a coherent state with 
$\langle x\rangle = 3\Delta x$ and $\langle p\rangle = 0$, so that 
the initial phase is zero. We see that the phase quickly drifts
to one of the two values $\pm\pi/2$, and from then on exhibits jumps in
the motion between these values. In Figure~\ref{fig1}(b) we show the position of the resonator as a function of time. The amplitude of the position oscillations tends to reduce during phase flips, as expected. 

One of the most interesting aspects of these quantum jumps is that
they are an emergent phenomenon; the underlying dynamics does not
contain jumps but consists purely of continuous diffusion.  The
authors of~\cite{Mabuchi98} explain this emergence by providing a
detailed analysis of the interplay of the correlations produced by the
Hamiltonian dynamics, the measurement induced localization, and the
diffusive noise. Their analysis is certainly not wrong. Our point here 
is that the mechanism that produces jumps in this case is the same mechanism 
that produces jumps in the case of a continuous QND measurement. This 
provides us with a somewhat simpler picture of the cause of the 
``retroactive'' jumps. 

QND measurements are measurements in which the observable being
measured is not changed by the dynamics of the system (that is, the
observable commutes with the Hamiltonian)~\cite{WMqoptics}.  As a
result, once the measurement has projected the system onto a given
eigenstate of the observable, it remains there throughout the
remainder of the observation period. Now consider what happens when
the observable is additionally subject to diffusion from environmental
noise. In the absence of the measurement the noise will cause the
observable to diffuse from one eigenstate to another, but in the
presence of a sufficiently strong continuous QND measurement the
dynamics is quite different. Consider what happens during a small time
interval when the system begins in one eigenstate. During the interval
the diffusion will generate a small probability that the system is in
an adjacent eigenstate. However, during the same interval the
measurement will collapse the system to one of the eigenstates, and
with very high probability this will be the initial eigenstate, since
the diffusion has only managed to generate a small probability for the
other eigenstates during the short time interval. As a result, it is
only unlikely events, in which the noise has a particularly large
fluctuation, and the measurement conspires by returning a low
probability result, that will cause the system to transition from one
eigenstate to the next. The result is periods in which the system
remains in a given eigenstate of the QND observable, interspersed by
quantum jumps between the eigenstates. The jumps are {\em quantum}
jumps since they only appear because the observable has a discrete
spectrum. This qualitative picture has been confirmed by numerical
simulations of a measurement of the energy of a quantum harmonic
oscillator~\cite{Santamore04,Jacobs07b}, and the states of an electron in a
coupled pair of quantum dots (often referred to as a ``charge
qubit'')~\cite{Goan01}.

The jumps discovered in~\cite{Mabuchi98}, and those in the dynamics
here, can now be seen to be caused by the same effect, except in that in
these two cases the QND measurement is mediated through a second
system.  In our case the QND observable is the energy of the CPB. The
interaction between the resonator and the CPB causes the phase of the
resonator to be tightly correlated with the energy eigenstates of the
CPB, as shown in Eq.(\ref{ssx}). In continually providing information
about the phase of the resonator, the position measurement necessarily
provides information about the energy of the CPB, generating a QND
measurement, and resulting in quantum jumps. Interestingly it is not 
necessary to perform a QND measurement on the mediating system (the
resonator) to generate the jumps; position is not a QND observable for
the resonator. 

We now consider the use of feedback control to stabilize the system 
against the jumps. In doing so we reveal a fundamental limitation of 
feedback control in this context - Hamiltonian feedback cannot stop 
the system from jumping. The reason for this is quite simple --- as 
discussed above, the jumps from $|-\rangle$ to $|\!+\rangle$, for 
example, are due to the small probability for $|\!+\rangle$ continually 
generated by the noise process. To generate this probability the noise 
process merely reduces the length of the Bloch vector. Since the cause 
of the jumps is purely this reduction (along with the QND measurement), 
and since Hamiltonian evolution, feedback or otherwise, can rotate 
the Bloch vector but cannot change its length, feedback is powerless 
to prevent the jumping. However, feedback can be used to {\em increase} 
the rate of jumps from either state, thereby altering the relative time 
that the system spends in the two states. Ultimately this could be used 
to ensure that the system spends almost all its time in one of the states, 
creating an effective stability. 

We implement this procedure by applying a feedback Hamiltonian of the form 
$H = \mu(n_x(t)\sigma_x + n_z(t)\sigma_z)$ to the CPB, where 
$n_x ^2+ n_z^2=1$, so that $\mu$ determines the feedback strength. In each 
time step $\Delta t =1/(2500f)$, we choose this Hamiltonian so as to rotate the 
system towards the state $|-\rangle$. We set $\mu=200$, and find that 
$\eta_{\mbox{\scriptsize tot}}=0.95$ is required to obtain a significant effect. 
We plot the resulting evolution in Fig~\ref{fig1}(c), for the same noise realization 
as before. This shows that the phase is now more stable at 
$-\pi/2$ than $\pi/2$.

\section*{Acknowledgments} KJ would like to thank A. N. Korotkov for helpful discussions. This work was supported by The Hearne Institute for Theoretical Physics, The Army Research Office and The Disruptive Technologies Office.

\vspace{0.7cm}

%\bibliographystyle{unsrt}
%\bibliography{report}

\begin{thebibliography}{10}

\bibitem{Mabuchi98}
H.~Mabuchi and H.~M. Wiseman.
\newblock Retroactive quantum jumps in a strongly coupled atom-field system.
\newblock {\em {Phys.\ Rev.\ Lett.}}, 81:4620, 1998.

\bibitem{Goan01}
H-S Goan and Gerard~J. Milburn.
\newblock Dynamics of a mesoscopic qubit under continuous quantum measurement.
\newblock {\em Phys. Rev. B}, 64:235307, 2001.

\bibitem{Santamore04}
D.~H. Santamore, A.~C. Doherty, and M.~C. Cross.
\newblock Quantum nondemolition measurement of fock states of mesoscopic
  mechanical oscillators.
\newblock {\em Phys. Rev. B}, 70(14):144301, 2004.

\bibitem{Santamore04b}
D.~H. Santamore, Hsi-Sheng Goan, G.~J. Milburn, and M.~L. Roukes.
\newblock Anharmonic effects on a phonon number measurement of a quantum
  mesoscopic mechanical oscillator.
\newblock {\em Phys. Rev. A}, 70:052105, 2004.

\bibitem{Martin07}
I.~Martin and W.~H. Zurek.
\newblock Measurement of energy eigenstates by a slow detector.
\newblock {\em Phys. Rev. Lett.}, 98:120401, 2007.

\bibitem{Jacobs07b}
K.~Jacobs, P.~Lougovski, and M.~P. Blencowe.
\newblock Continuous measurement of the energy eigenstates of a nanomechanical
  resonator without a nondemolition probe.
\newblock {\em Phys. Rev. Lett.}, 98:147201, 2007.

\bibitem{Buks06}
Eyal Buks, Eran Arbel-Segev, Stav Zaitsev, Baleegh Abdo, and M.~P. Blencowe.
\newblock Quantum nondemolition measurement of discrete fock states of a
  nanomechanical resonator.
\newblock Eprint: arXiv:quant-ph/0610158, 2006.

\bibitem{Jacobs08}
K.~Jacobs, A.~N. Jordan, and E.~K. Irish.
\newblock Energy measurements and preparation of canonical phase states of a
  nanomechanical resonator.
\newblock Eprint: arXiv:0707.3803, 2007.

\bibitem{Makhlin01}
Y.~Makhlin, G.~Sch\"{o}n, and A.~Shnirman.
\newblock Quantum-state engineering with josephson-junction devices.
\newblock {\em Rev. Mod. Phys.}, 73:357, 2001.

\bibitem{Naik06}
A.~Naik, O.~Buu, M.~D. LaHaye, A.~D. Armour, A.~A. Clerk, M.~P. Blencowe, and
  K.~C. Schwab.
\newblock Cooling a nanomechanical resonator with quantum back-action.
\newblock {\em Nature}, 443:193, 2006.

\bibitem{Korotkov}
Alexander~N. Korotkov.
\newblock Selective quantum evolution of a qubit state due to continuous
  measurement.
\newblock {\em Phys. Rev. B}, 63:115403, 2001.

\bibitem{Armour02}
A.~D. Armour, M.~P. Blencowe, and K.~C. Schwab.
\newblock Entanglement and decoherence of a micromechanical resonator via
  coupling to a cooper-pair box.
\newblock {\em {Phys.\ Rev.\ Lett.}}, 88:148301, 2002.

\bibitem{Hopkins03}
Asa Hopkins, Kurt Jacobs, Salman Habib, and Keith Schwab.
\newblock Feedback cooling of a nanomechanical resonator.
\newblock {\em Phys. Rev. B}, 68:235328, 2003.

\bibitem{JacobsSteck06}
K.~Jacobs and D.~Steck.
\newblock A straightforward introduction to continuous quantum measurement.
\newblock {\em Contemporary Physics}, 47:279, 2006.

\bibitem{Brun02}
Todd~A. Brun.
\newblock A simple model of quantum trajectories.
\newblock {\em {Am.\ J.\ Phys.}}, 70:719, 2002.

\bibitem{WienerIntroPaper}
D.~T. Gillespie.
\newblock The mathematics of brownian motion and johnson noise.
\newblock {\em Am. J. Phys.}, 64:225, 1996.

\bibitem{DJ}
A.~C. Doherty and K.~Jacobs.
\newblock Feedback control of quantum systems using continuous state
  estimation.
\newblock {\em Phys.\ Rev.\ A}, 60:2700, 1999.

\bibitem{Ruskov05}
Rusko Ruskov, Keith Schwab, and Alexander~N. Korotkov.
\newblock Squeezing of a nanomechanical resonator by quantum nondemolition
  measurement and feedback.
\newblock {\em Phys. Rev. B}, 71:235407, 2005.

\bibitem{Caldiera89}
A.O. Caldeira, H.A. Cerdeira, and R.~Ramaswamy.
\newblock Limits of weak damping of a quantum harmonic oscillator.
\newblock {\em Phys. Rev. A}, 40:3438, 1989.

\bibitem{Clerk04}
A.~A. Clerk, S.M. Girvin, A.~K. Nguyen, and A.~D. Stone.
\newblock Resonant cooper-pair tunneling: Quantum noise and measurement
  characteristics.
\newblock {\em {Phys.\ Rev.\ Lett.}}, 89:176804, 2002.

\bibitem{WMqoptics}
D.~F. Walls and G.~J. Milburn.
\newblock {\em Quantum Optics}.
\newblock Springer, New York, 1995.

\end{thebibliography}

\end{document}